\documentclass{article}

\newtheorem{theorem}{Theorem}[section]

\newtheorem{proposition}[theorem]{Proposition}

\newtheorem{lemma}[theorem]{Lemma}

\newtheorem{example}[theorem]{Example}

\newtheorem{claim}[theorem]{Claim}

\renewcommand{\Box}{{\vrule width0.6ex height1em depth0cm}}

\newenvironment{proof}{\noindent{\bf Proof:}}{\hfill \Box}

\newenvironment{newproof}[1]{\noindent{\bf Proof of #1:}\

}{$\Box$\medskip\par}

\def\build#1_#2^#3{\mathrel{\mathop{\kern 0 pt#1}\limits_{#2}^{#3}}}

\usepackage{graphicx}
\usepackage{tabularx}
\usepackage{float}

\newcommand{\ra}{\rightarrow}
\newcommand{\lra}{\longrightarrow}
\newcommand{\calM}{M}

\newcommand{\kxorsat}{k\hbox{-XOR-SAT}}

\begin{document}

\title{Approximating  the satisfiability threshold for random $k$-XOR-formulas}

\author{Nadia CREIGNOU\thanks{ E-mail:
creignou@lim.univ-mrs.fr }   \\
         Laboratoire d'Informatique de Marseille\\
         LIM, FRE CNRS 2246\\
         Universit\'e de la M\'editerran\'ee\\
Marseille, France
                \and
        Herv\'e DAUD\'E \thanks{ E-mail:
daude@gyptis.univ-mrs.fr }  \\
Centre de
Math\'ematiques et d'Informatique\\
                LATP, UMR CNRS 6632\\
        Universit\'e de Provence\\
Marseille, France
 \and
 Olivier DUBOIS\thanks{ E-mail:
Olivier.Dubois@lip6.fr }  \\
CNRS-Universit\'e de Paris 6\\
LIP6\\
Paris, France
        }

\date{\today}

\bibliographystyle{alpha}

\maketitle

\begin{abstract}
In this paper we study random linear systems with $k$ variables
per equation over the finite field $GF(2)$, or equivalently
$k$-XOR-CNF formulas. In a previous paper Creignou and Daud\'e
proved that the phase transition for the consistency
(satisfiability) of such systems (formulas) exhibits a sharp
threshold. Here we prove that the phase transition occurs as  the
number of equations (clauses) is proportional to the number of
variables. For any $k\ge 3$ we establish first estimates for the
critical ratio. For $k=3$ we get $0.93$ as an upper bound, $0.89$
as a lower bound, whereas experiments suggest that the critical
ratio is approximately $0.92$.
\end{abstract}

\section{Introduction}

For $k$-CNF formulas many experiments have shown a very swift
transition between satisfiability and unsatisfiability as the
ratio $L/n$ of the number of clauses over the number of variables
is varied. That is, there exists $c_k$, a critical value of $L/n$
, such that if $L/n<c_k$ then the formula is almost surely
satisfiable and if if $L/n>c_k$ then the formula is almost surely
unsatisfiable. Most of the papers investigating this phase
transition are directed towards obtaining approximate estimates of
its location. For instance for $3$-SAT, for an observed sharp
threshold of about $c_3=4.25$ the best lower bound is $3.003$
\cite{FriezeS-96} and  tight  upper bounds have been successively
obtained,  4.601 \cite{KiroukisKKS-98}, 4.596 \cite{JansonSV-00}
and finally  4.506 \cite{DuboisBM-00}. In 1999, Friedgut proposed
a new and fruitful approach. In a remarkable paper
\cite{Friedgut-99}, with an appendix by Bourgain, he developed a
general sharp threshold criterion for monotone subsets of the
hypercube. In using this criterion he proved that $k$-SAT
\cite{Friedgut-99} and (in collaboration with Achlioptas) $k$-col
\cite{AchlioptasF-99} exhibit a sharp threshold. In
\cite{CreignouD-00} Creignou and Daud\'e applied this criterion to
the $k$-XOR-SAT problem (in which the usual ``or" is replaced by
the ``exclusive or"). Thus, they proved the existence of a sharp
threshold phenomenon for $k$-XOR-SAT, $k\ge 3$, without specifying
its location.

The aim of this paper is to prove that the phase transition for
$k$-XOR-SAT occurs as the number of clauses is proportional to the
number of variables and more precisely to provide approximate
estimates of its location. In Section \ref{lower bound section} we
give a lower bound and in  Section \ref{upper bound section} an
upper bound. Our results rely first on the sharpness of the
threshold established in \cite{CreignouD-00} (see Section
\ref{definitions section}) and second on the handy translation of
our problem in terms of random matrices (see Section \ref{random
matrices section}). These theoretical results are supplemented
with  experiments in Section \ref{experiments section}.

\section{Notation and definitions}\label{definitions section}

Throughout the paper $k$ will denote an integer equal to or greater than $3$.
A {\it
$k$-XOR-clause} (or shortly a {\it $k$-equation}), $C$,  is
a linear equation over the finite field $GF(2)$ using exactly $k$ variables,
$C=((x_1\oplus\ldots \oplus x_k)= \varepsilon)$ where $\varepsilon = 0\hbox
{ or
} 1$.
A {\it $k$-XOR-formula}
(or shortly a {\it $k$-system})
is a conjunction of
distinct  $k$-XOR-clauses. A {\it truth assignment}
$I$ is a mapping
that assigns $0$ or $1$ to each variable in its domain, it satisfies
an XOR-clause
  $C=((x_1\oplus\ldots \oplus
x_k)= \varepsilon)$\  iff \
$\displaystyle I(C):=\sum _{i=1}^p I(x_i)\bmod 2=\varepsilon$ and it
satisfies a formula $F$
iff it satisfies every clause in $F$. \\
We will denote by $\kxorsat$ (or shortly SAT) the property for a
$k$-XOR-formula of being satisfiable (or equivalently the property
for a $k$-system of being consistent)
  and by UNSAT the property  of being
unsatisfiable.
The property UNSAT is monotone increasing.\\
Throughout the paper we reserve $n$ for the number of variables
($\{x_1,\ldots ,x_n\}$
denotes the set of variables). There are $N_k=2 {n\choose k}$ different
$k$-XOR-clauses over $n$ variables.
We consider the random formula obtained by choosing uniformly,
independently and with replacement $L$
clauses
from the $N_k$ possible $k$-clauses. This defines a probability space
of $k$-XOR-formulas consisting
of all ordered sets of $L$  clauses, not necessarily distinct, with
$k$ distinct variables, each over
a set of Boolean variables, each set of clauses having the same
probability. This probability space
is denoted by $\Omega(n, L, k)$, the associated probability is the uniform law:

$$\forall s  \in  \Omega(n, L, k) \quad  P_{n,L} (s)= \Biggl( 2{n
\choose k}\Biggr)^{-L}$$
We are interested in estimating the probability that a formula drawn
at random from  $\Omega(n, L,
k)$
  is satisfiable, that is in estimating
$P_{n,L}(\kxorsat)$.

Our model for producing a random formula relates to the model in which
each of the possible clauses is chosen independently with probability $p$
  in the same way $G(n,L)$ relates to $G(n,p)$ in random graph theory.
In most investigations on
properties of random subgraphs of the complete graph on $n$ vertices,
these two models are practically
interchangeable, provided the number, $L$, of edges be close to $p
{n\choose 2}$ (see \cite{Bollobas-85}).
In \cite{CreignouD-00} the first  two authors proved that when each  clauses is chosen
with probability $p$ then $\kxorsat$ exhibits a sharp
threshold for $k\ge 3$. They also noted that the transition occurs
when $ pN_k \leq n$. Thus in the
context of random formula   these results are still valid in our
model ( provided  $p\sim L/N_k$), they
can be expressed as:

\begin{theorem}{\rm \cite{CreignouD-00}}\label{sharp threshold thm}

For every  $k\ge 3$ there exists a function $c_k(n)\le 1$ such that:
for every  $\varepsilon>0$

$\displaystyle \lim_{n\rightarrow +\infty}
P_{n,(c_k(n)-\varepsilon)n}(\kxorsat)=1$

$\displaystyle\lim_{n\rightarrow +\infty}
P_{n,(c_k(n)+\varepsilon)n}(\kxorsat)=0 $
\end{theorem}

We are going to prove that this sharp threshold behavior occurs as
the number
  of clauses is proportional to
the number of variables of formulas and more precisely we are going
to provide lower and upper bounds for $c_k(n)$.
Since we know from Theorem \ref{sharp threshold thm} that $c_k(n)\le
1$, we can and we will
suppose in the sequel that $L\le n$. \\
As mentioned in \cite{Friedgut-99}, though there is a swift
transition of probability
  of satisfiability it is still feasible that the critical value $c_k(n)$
does not converge to any given value.
However, one can define $\displaystyle\beta_k =\limsup _{n\ra +\infty} c_k(n)$
and $\displaystyle\alpha_k =\liminf _{n\ra +\infty} c_k(n)$.
  Thus, $\beta_k$ (respectively, $\alpha_k$)
is the least (greatest) real number such that if $c>\beta_k$
($c<\alpha_k$) then the probability of a $k$-XOR-formula with $n$
variables and $cn$ clauses being satisfiable converges to $0$ (to
$1$) as $n$ tends to infinity; hence $\alpha_k\le \beta_k$ and
  experiments suggest strongly (see Section \ref{experiments section})
that equality holds.
\medskip

In a first step (Section \ref{random matrices section}) we will reduce our
problem to the study of the rank of random sparse matrices over $GF(2)$.
Then, we will give a lower bound (Section
\ref{lower bound section}) and
an  upper bound (Section \ref{upper bound section}) for the threshold.

At last, $H(x)$  will denote the well-known entropy function,
$$H(x)= x\ln(x)+(1-x)\ln(1-x).$$

\section{From random formulas to random matrices}\label{random
matrices section}

The aim of this section is to express $P_{n,L}(\kxorsat)$ in terms of
random matrices. We will first
use  a technique related to the harmonic mean formula  and
publicized in \cite{KamathMPS-94} and
\cite{Aldous-89}. This technique provides a simple expression for the
number of satisfiable systems and
can be described  by a bipartite graph $G$ formed by $\{0,1\}^n$ (the
set of all  assignments $I:\
\{x_1,\ldots ,x_n\} \lra \{0,1\} $) as the first part of vertices and
$\Omega(n, L, k)$ (the set of all systems s) as the second part. There is an
  edge $\{I,s\}$ in $G$  if the assignment $I$ satisfies the system $s$.
  The degree of every assignment $I$  is  equal to $ {n\choose k}^L$ and
for every  system $s$  the degree $d(s)$  of $s$  is the number of distinct
assignments satisfying $s$. The number of satisfiable $k$-systems is
thus given by
$$\vert \kxorsat\vert=\sum_{I\in  \{0,1\}^n}\ \ \ \sum_{s, I(s)=1}
{1\over d(s)}.$$ As there is a one-to-one   correspondence from
the set of linear systems satisfied by some fixed assignment $I_0$
onto the set $\cal H$ of homogeneous linear systems, we get:
$$\vert \kxorsat\vert=2^n\sum_{s\in \cal H}
{1\over d(s)}.$$
 From this simple expression  we get the following theorem.

\begin{theorem}\label{matrix theorem}
  Let $Y(A)$ denote the number of vectors in the kernel of the transpose
of $A$,  then providing $A$ be chosen uniformly in the set  $\calM
_{L,n,k}$ of Boolean $L\times n$ matrices
with exactly $k$ units  in each row:
$$P_{n,L}(\kxorsat)=\sum_{r=0}^L 2^{r-L}P_{(r)}=E_{L,k}(1/Y)$$
where $P_{(r)}$ is the probability that a matrix from $
\calM_{L,n,k}$ is of rank $r$,
and $ E_{L,k}$ denotes  the expectation.

\end{theorem}

\begin{proof}
 From our preliminary work we have:
$$\vert \kxorsat \vert=2^n\sum_{A\in \calM_{L,n,k}}
{1\over 2^{n-rank  (A)}}
=
\sum_{r=0}^L
\sum_{A\in \calM_{L,n,k}\atop rank (A)=r}
2^r.$$

But  $\#\{A\in \calM_{L,n,k}\ /\
rank (A)=r\} =\vert \calM_{L,n,k}\vert  P_{(r)} $
and

  $P_{n,L}(\kxorsat)=\vert \kxorsat \vert/  \left ( 2{n \choose
k}\right )^L$, as $\vert \calM_{L,n,k}\vert
={n\choose k}^L $ we get:
$$P_{n,L}(\kxorsat)=\sum_{r=0}^L 2^{r-L}P_{(r)}.$$

The second equality is justified by the  following fact:

  $$( rank  (A)=r)  \Longleftrightarrow Y= 2^{L-r}.$$

\end{proof}
\medskip

We will see that this result leads to two key facts, Proposition
\ref{key fact 1} and Proposition \ref{matrix corollary}, which
enable us to give tight bounds for the threshold.

\section{A lower bound for the threshold}\label{lower bound section}

By  Theorem \ref{matrix theorem} and
   Jensen's inequality we have:
\begin{equation}\label{E(Y) equation}
P_{n,L}(\kxorsat)=E_{L,k}(1/Y)\ge{1\over E_{L,k}(Y)}.
\end{equation}
Thus, we get the first key fact:

\begin{proposition} \label{key fact 1}
$\displaystyle \hbox { If } \lim_n E_{cn,k}(Y)=1 \hbox { then } \alpha_k \geq c.  $

\end{proposition}

A first lower bound for the threshold of $k$-XOR-SAT  can  be derived
from the following result
\begin{theorem}\label{lower bound thm}
Let $\theta_k$ denote
the minimum  over $[0,1/2]$ of the function  $$f_k(x)= {\ln2+H(x)
\over \ln(1+(1-2x)^k)}$$
then  for any $\displaystyle c<\theta_k, \quad  \lim_n  E_{cn,k} (Y)=1$

\end{theorem}

From MAPLE's estimates for $\theta_k$  we get the following
bounds:
\begin{equation}\label{lower bound}
{\bf \alpha_3\ >\ 0.88949} ,\quad {\bf \alpha_4\ >\ 0.96714}
,\quad{\bf \alpha_5\ >\ 0.98916} ,
\quad{\bf
\alpha_6\ >\ 0.99622}
\end{equation}

To prove Theorem \ref{lower bound thm}
we  first   show that the behavior of $E_{L,k}(Y)$ is given by the one
of the following quantity:

  $$  \omega_{n,m,k} =\displaystyle \sum_{s=0}^k (-1)^s{m\choose
s}{n-m\choose k-s}.$$

\begin{proposition}\label{E(Y) lemma}
$\displaystyle E_{L,k}(Y)=2^{-n}\sum_{m=0}^n {n\choose
m}\Bigl(1+{\omega_{n,m,k}\over {n\choose k}}\Bigr)^L .$
\end{proposition}

\begin{proof}
Let us consider  $X(A)=\vert \ker A\vert$,
  $X$ is the sum of indicator variables:
$$X=\sum_{\vec u\in\{0,1\}^n} X_{\vec u}\hbox { where }
X_{\vec u}(A)=1\hbox { if and only if } \vec u\in \ker \! A.$$
By symmetry, $\displaystyle E(X)=\sum_{m=0}^n {n\choose m}E(X_{\vec u_m})$
  where $\vec u_m$ is any vector of weight $m$.\\
For each $\vec u_m $, consider the subset $I_{ \vec u_m}$ of
$\{1,\cdots,n\}$   formed by the indexes of the unit coordinates
of $\vec u_m$. Then $\vec u_m$    is in the kernel of $A$ if and
only if  each row of the submatrix of $A$ formed by the $m$
columns whose indexes are in $I_{ \vec u_m}$ has an even  number
$s$ of unit coefficients. Thus, we have
$$ \sum_{s=0\bmod 2} {m\choose s}{n-m\choose k-s} $$
possible rows for a matrix $A$ in $M_{n,L,k}$ such that
$X_{\vec u_m}(A)=1$, and we deduce:

$$ {n\choose k}^L\ E(X_{\vec u_m})=\left ( \displaystyle
\sum_{s=0\bmod 2} {m\choose s}{n-m\choose k-s}
\right ) ^L= \Bigl ( {{n\choose k} + \omega_{n,m} \over 2} \Bigr ) ^ L,$$
therefore,
$\displaystyle E(X)=2^{-L}\sum_{m=0}^n {n\choose
m}(1+{\omega_{n,m,k}\over {n\choose k}})^L.$

As $Y=2^{L-n}X$, the conclusion follows.
  \end{proof}

\medskip

This result shows that if we  split $\displaystyle E_{L,k}(Y) $ in
two   then the  proof of Theorem \ref{lower bound thm} follows
from the two following claims:

\begin{claim}\label{claim1} If $L\leq n$ then $\displaystyle\lim_{n}\Sigma_1(L,k) =1$, where
$$\Sigma_1(L,k)  =2^{-n}\sum_{\vert m-{n\over 2}\vert <n^{1-{4 \over 3k}}}
{n\choose m}(1+{\omega_{n,m,k}\over {n\choose k}})^L. $$

\end{claim}

\begin{claim}\label{claim2} If $c<\theta_k $ then
$\displaystyle\lim_{n}\Sigma_2(\theta n,k) =0$, where
$$\Sigma_2(L,k)  =2^{-n}\sum_{\vert m-{n\over 2}\vert\ge n^{1-{4 \over 3k}}}
{n\choose m}(1+{\omega_{n,m,k}\over {n\choose k}})^L.$$
\end{claim}

These claims rely on asymptotical properties of sums of binomial
coefficients and of the quantity
$\omega_{n,m,k}$. Observe that for each $k$,  $\omega_{n,m,k}$ can be
viewed as a polynomial over two variables $n$ and
$m$ of total degree $k$, this is made precise by the following:

\begin{proposition}\label{omegas behavior lemma}
$$\omega_{n,m,k}= {(n-2m)^k\over k!}-  {n(n-2m)^{k-2}\over 2(k-2)!} +
P_k(n,m)$$
where $P_k$ is a polynomial on two variables  of  total degree $k-2$. Thus,
there exists absolute positive constants $A_k$ and $B_k$ such that:

\begin {equation} \label {key for w}
  \Bigl \vert {\omega_{n,m,k} \over {n\choose k}} - (1- {2m \over n}
)^k \Bigr \vert  \leq {A_k \over n }
\Bigl \vert 1- {2m \over n}  \Bigr \vert ^{k-2} + {B_k \over n^2}
\end {equation}
\end{proposition}

\begin{proof}
Let us show how we get the leading term.
Observe that:
$$\omega_{n,m,k}=\sum_{s=0}^k (-1)^s{m^s \over s!}{(n-m)^{k-s} \over (k-s)!}+P(n,m),$$
where $P(n,m)$ is a polynomial of degree $k-1$. But,
$$\sum_{s=0}^k (-1)^s{m^s \over s!}{(n-m)^{k-s} \over (k-s)!}
={1 \over k!}\sum_{s=0}^k {k\choose s} (-m)^s (n-m)^{k-s}
={1 \over k!}(n-2m)^k.$$
The study of the second term is left to the reader.
\end{proof}
\medskip

>From this result and from its following consequences, Lemma \ref{central area lemma}
and Lemma \ref{extremal area lemma}, it turns out that that the asymptotical behavior of
$\Sigma_1(L,k) $ and $\Sigma_2(L,k) $ is given by
the one of $\displaystyle (1+(1- {2m \over n} )^k)^L$.
%Indeed  we will
%need the two following   asymptotic
%consequences in which  constants in the big O depend only on $k$.

\begin{lemma}\label{central area lemma}  If $L\leq n$ and $ \vert
m-{n\over 2}\vert <n^{1-{4 \over 3k}}$ then
$$\Bigl \vert \bigl ( 1+ {\omega_{n,m,k} \over {n\choose k}} \bigr ) ^L -
\bigl ( 1+(1- {2m
\over n} )^k \bigr)^L \Bigr \vert = O(n^{-7/9}) ,$$
$$\Bigl \vert
\bigl ( 1+(1- {2m
\over n} )^k \bigr )^L - 1  \Bigr \vert = O(n^{-1/3}).$$
  \end{lemma}

\begin {proof}
From (\ref {key for w}) and from the mean value theorem we have  when
$L \leq n$:
$$ \Bigl \vert \bigl ( 1+ {\omega_{n,m,k} \over {n\choose k}} \bigr ) ^L -
\bigl ( 1+(1- {2m
\over n} )^k \bigr)^L \Bigr \vert \leq \Biggl ( {A_k }
\Bigl \vert 1- {2m \over n}  \Bigr \vert ^{k-2} + {B_k \over n}
\biggr ) \Bigl (1 + \gamma_{n,m,k}\Bigr )^{L-1}
$$
with $\gamma_{n,m,k} $ lying between  $  {\omega_{n,m,k} \over
{n\choose k}}$ and $ (1- {2m
\over n} )^k $.

As   $ \vert m-{n\over 2}\vert <n^{1-{4 \over 3k}}$  and $k\geq 3$ we
get
  $\displaystyle  \vert 1-{2m\over n}\vert^{k-2} =O (n^{-4 \over 9})$. Thus,
  (\ref {key for w}) gives
$ \displaystyle   \Bigl \vert {\omega_{n,m,k} \over {n\choose k}} -
(1- {2m \over n} )^k \Bigr \vert =O(n^{-13 \over
9}) $ which shows that $\displaystyle {\omega_{n,m,k}
\over {n\choose k}}= O(n^{-4\over 3})$. Since
we also have $\displaystyle  \vert 1-{2m\over n}\vert^k =
O(n^{-4 \over 3})$, we can conclude that
$ \gamma_{n,m,k} $ are  $ O(n^{-4
\over 3})$.  This leads to $\Bigl (1 + \gamma_{n,m,k}\Bigr )^{L-1} =
O(n^{-1 \over 3})$ and  the rest of the proof
is now pure routine.
\end {proof}
\medskip

\begin{lemma}\label{extremal area lemma}  If $L\leq n$,  and
$m-{n\over 2} \leq - n^{1-{4 \over 3k}}$
or $k$ even and $m-{n\over 2} \geq  n^{1-{4 \over 3k}}$
then
$$\Biggl \vert {\bigl ( 1+ {\omega_{n,m,k} \over {n\choose k}} \bigr ) ^L \over
\bigl ( 1+(1- {2m
\over n} )^k \bigr)^L}  \Biggr   \vert = O(1) .$$
  \end{lemma}

\begin {proof}
From (\ref {key for w}) and from the mean value theorem we get:
$$  \Bigl \vert \ln \bigl ( 1+ {\omega_{n,m,k} \over {n\choose k}} \bigr ) -
\ln  \bigl ( 1+(1- {2m
\over n} )^k  \bigr )  \Bigr \vert \leq  \Biggl ( {A_k\over n }
\Bigl \vert 1- {2m \over n}  \Bigr \vert ^{k-2} + {B_k \over n^2}
\biggr ) \Bigl ({ 1\over  1+ \nu_{n,m,k}}\Bigr )
$$
with $\nu_{n,m,k} $ lying between  $  {\omega_{n,m,k} \over {n\choose
k}}$ and $ (1- {2m
\over n} )^k $.\\
Under the lemma's assumptions we get
$\displaystyle ( 1-{2m\over n})^{k-2} \geq
2^{k-2}n^{-4(k-2) \over 3k}$.\\
Thus,   (\ref {key for w}) gives:
$ \displaystyle   {\omega_{n,m,k} \over {n\choose k}}\geq
2^{k-2}n^{-4(k-2) \over 3k} \Bigl [ 4 n^{{-8 \over 3k}} - A_k
  n^{-1}  \Bigr ] - B_k  n^{-2}$\\
  which shows that there exits $C_k$ such that for sufficiently large $n$

$\displaystyle {\omega_{n,m,k} \over {n\choose k}} \geq C_k n^{-4/3}>0. $
As $ (1- {2m\over n} )^k >O $ we get   $ \nu_{n,m,k} >0 $ and   with
$L \leq n$, the first
inequality leads to:
$ \displaystyle \Biggl \vert {\bigl ( 1+ {\omega_{n,m,k} \over
{n\choose k}} \bigr ) ^L \over
\bigl ( 1+(1- {2m
\over n} )^k \bigr)^L}  \Biggr   \vert\leq  \exp \Bigl (  A_k + {B_k
\over n}\Bigr ).$
\end {proof}

\bigskip

As mentioned above, the proof of Claims  \ref{claim1} and
\ref{claim2} will follow from  well-known results on the behavior
of the binomial coefficients and on the distribution of the
binomial law.\medskip

\begin{newproof}{Claim  \ref{claim1}}
As $\displaystyle 2^{-n}\sum_{ m=0}^n {n\choose m}=1$, Lemma (\ref
{central area lemma}) implies that:
$$ \Bigl \vert \Sigma_1(L,k)  - 2^{-n}\sum_{\vert m-{n\over 2}\vert<
n^{1-{4 \over 3k}}} {n\choose m} \Bigr \vert =   O(n^{-1/3}).$$
But, $k\geq 3$ hence $1-{4 \over 3k}>{1 \over 2}$ and  De
Moivre-Laplace's theorem asserts that:
$$\lim_{n\ra +\infty} 2^{-n}\sum_{\vert m-{n\over 2}\vert <n^{1-{4 \over 3k}}}
{n\choose m}  =1.$$
\end{newproof}

\begin{newproof}{Claim  \ref{claim2}}
On the one hand, when $k$ is odd and  $ m-{n\over 2} \ge n^{1-{4
\over 3k}}$, $-1 \leq 1+(1- {2m \over n} )^k \leq 2^{-k}n^{-4
\over 3} $. Similar computations as in the proof of Lemma
\ref{extremal area lemma} show that for sufficiently large $n$,
$\displaystyle -1\leq {\omega_{n,m,k} \over {n\choose k}} <0$,
thus
$\displaystyle 0 \leq \bigl ( 1+ {\omega_{n,m,k} \over {n\choose k}} \bigr ) ^L  \leq 1 $.\\
From   De Moivre-Laplace's theorem we know that  for any  $k\geq 3:$
$$\lim_{n\ra +\infty}2^{-n}\sum_{\vert m-{n\over 2}\vert \geq n^{1-{4
\over 3k}}}
{n\choose m} =0.$$
Therefore, when $k$ is odd, following  Lemma \ref{extremal area lemma}
$$\Sigma_2(L,k)=
\sum_{ m-{n\over 2} \leq -n^{1-{4  \over 3k}}}
{n\choose m}(1+(1- {2m \over n} )^k)^L +o(1).$$

On the other hand, when $k$ is even  Lemma \ref{extremal area lemma}
 shows that
$$\Sigma_2(L,k)=O
\bigl (
\sum_{\vert m-{n\over 2}\vert \geq n^{1-{4 \over 3k}}}
{n\choose m}(1+(1- {2m \over n} )^k)^L
\bigr ).$$
By parity and symmetry of binomial coefficients
$$\sum_{ m-{n\over 2} \geq n^{1-{4  \over 3k}}}
{n\choose m}(1+(1- {2m \over n} )^k)^L = \sum_{ m-{n\over 2} \leq
-n^{1-{4  \over 3k}}} {n\choose m}(1+(1- {2m \over n} )^k)^L.$$
Therefore,
 when $k$ is even
$$\Sigma_2(L,k)=O
\bigl (
\sum_{ m-{n\over 2} \leq -n^{1-{4  \over 3k}}}
{n\choose m}(1+(1- {2m \over n} )^k)^L
\bigr ).$$
Therefore, Claim
\ref {claim2} is proved as soon as we  are able to prove that  for any
positive real  $c<\theta_k:  $

\begin{equation}\label{avantlast}
  T_{c,n} =\  2^{-n} \sum_{  m-{n\over 2} \le - n^{1-{4 \over 3k}}}
{n\choose m}\bigl ( 1+(1- {2m
\over n} )^k \bigr)^{cn}=o(1).
\end{equation}

From Stirling's formula (see \cite{Bollobas-85}),  we have for any
$m\ge 1$,
$${n\choose m}
\le
{1\over \sqrt{2\pi}}\left ({n\over m(n-m)}\right )^{1/2} \exp \biggl
(-n\ H\,(\,{m\over n} \,)\biggr)
e^{1/12n};$$
thus,
$${n\choose m}=O
\biggl ( exp \biggl
(-n\ H\,(\,{m\over n} \,)\biggr)\biggr)
.$$
Therefore, \begin{equation}\label{last}
  T_{c,n} = 0 \Bigl ( \sum_{  m-{n\over 2} \le n^{1-{4 \over 3k}}}
\exp \bigl (-n \ g_{c,k}\,(\,{m\over n}\,) \bigr )
\Bigr )
\end{equation}
  where
$$ g_{c,k}(x)= \ln(1+(1-2x)^k)(f_k(x)-c) .$$
  Now let us  first note  that for any   $c < \theta_k$,
$(f_k(x)-c)\ge \theta_k-c>0$.
Hence, for any $\alpha >0$ there exists $\delta>0$ such that:
  $$g_{c,k}(x)>\delta  \hbox { when }  x \in [0,1/2-\alpha].$$
Then, the Taylor expansion  gives for any $k\geq 3:$
$g_{c,k}(1/2-t)= 2t^2 +o(t^2)$.  Thus, there exits an absolute positive
constant $C_k$ such that
   $$g_{c,k}(x)\ge C_k n^{-8/3k}  \hbox { when } x \in [1/2-\alpha,
1/2 -n^{-4/3k}].$$
This proves the existence of some $K>0$ such that  when   $m - n/2
\leq  -n^{1-4/3k}$,
$$\displaystyle \exp \bigl (-n \ g_{c,k}\,({m\over n}) \bigr ) \le
\exp(-Kn^{1/9}),$$
  with (\ref{last}), this establishes  (\ref{avantlast}) and thus
Claim \ref {claim2}.
\end{newproof}

\section{An upper bound for the threshold}\label{upper bound section}

Let  $Q$ denote the property ``$A$ is
of maximal rank" then
$Pr(Q)= P_{(L)}.$

Since $P_{(L)}+P_{(L-1)}\ldots +P_{(0)}=1$, we have
$\displaystyle\sum^L_{r=0} 2^{r-L} P_{(r)}\le  P_{(L)}+ {1
-P_{(L)}\over 2}$. Therefore, from  Theorem \ref{matrix theorem}
we get:
$$ P_{(L)}\le P_{n,L}(\kxorsat)\le {1 +P_{(L)}\over 2}.$$

  Since we know that $\kxorsat$ exhibits a sharp threshold
(see Theorem \ref{sharp threshold thm}) we get the  second key fact:
\begin{proposition}\label{matrix corollary}
  The decreasing property $Q$ exhibits a sharp threshold
  whose location coincides exactly with the one of $\kxorsat$.
Moreover  if  for some  $c$ there exist $\epsilon >0$ such that
for $L=cn$  and all (sufficiently large) $n$

$$   P_{(L)}<1- \epsilon \hbox {  then }  \beta_k < c.  $$
\end{proposition}

In other words,  $c$ is an upper bound for the threshold of $k$-XOR-SAT
as soon as we are able to prove that a  matrix from $M_{cn,n,k}$ is
not of maximal
rank with positive probability.
\medskip

When $k=3$, such a bound   can  be derived from the following result:
\begin{theorem}\label{upper bound thm}
$$P_{(L)}\le 1-{\exp (-3c)+3c\exp(-6c)-c\exp(-9c) -1+c\over c} +o(1).$$
\end{theorem}
Therefore, solving  the equation
$$\exp (-3c)+3c\exp(-6c)-c\exp(-9c) -1+c=0,$$
in using MAPLE provides

\begin{equation}\label{upper bound}
{\bf \beta_3\  <\  0.9278.}
\end{equation}
\medskip

\begin{proof}
The matrices we consider, in $M_{L,n,3}$, have three units by row.
Let $A=(a_{i,j})_{i=1,\cdots ,L\atop j=1,\cdots ,n}$ be such a
matrix. For each row $(a_{i,1},\ldots,a_{i,n})$, $i=1,\ldots ,L$,
of $A$ there are three distinct indices $1\le i_1<i_2<i_3\le n$
such that $a_{i,j} =1$ if $j=i_1, i_2$ or $i_3$, and $0$
otherwise; let $I_i(A)$ denote the set $\{i_1, i_2, i_3\}$. So, to
each row  we can associate its $3$-weight, $W_i(A)$ defined as the
following multiset:
$$W_i(A)=\big \{\sum_{k=1}^L a_{k,i_1}, \sum_{k=1}^L a_{k,i_2}, \sum_{k=1}^L a_{k,i_3}\big\}.$$

Now, let us introduce three random variables $T$, $U$ and $V$. \\
For any
random matrix $A$,
let $T(A)$ count the number
of all-zero columns in $A$.Observe that these columns do not contribute to
the rank of $A$.\\
Let $U(A)$ be defined by
$$U(A)=\#\{ i\ /\ W_i(A)=\{1,1,1\}\}.$$
Observe that each row contributing to $U(A)$ underlines three columns such that only one of them
contributes to
the rank of $A$.\\
Finally, let $V(A)$ be defined by
$$V(A)=\#\{ i\ /\ W_i(A)=\{1,1,\alpha\}\hbox{ with }\alpha\ge 2\}.$$
Observe that each row contributing to $V(A)$ underlines three columns such that two  of them
contribute to
the rank of $A$.\\
Thus, we have
$$Rank(A) \le Min(L,n-T(A)-2U(A)-V(A)).$$
Hence,
$$\hbox{if } T+2U+V>n-L\hbox{ then } A \hbox{ is not of maximal rank, }L .$$
Therefore,
\begin{equation}\label{uvw-equation}
P_{(L)}\le Pr(T+2U+V\le n-L).
\end{equation}
 Observe that
$$
E(T+2U+V)=\sum_{i=1}^n Pr(T+2U+V\ge i).
$$
Thus, for any integer $e\le n$,
$$
E(T+2U+V)\le e+(n-e)Pr(T+2U+V\ge e+1).
$$
Hence,
$$Pr(T+2U+V\ge e+1)\ge {E(T+2U+V)-e\over n-e}.$$
Thus (\ref{uvw-equation}) gives for $L=cn$
$$P_{(L)}\le 1 - {{E(T+2U+V)\over n} -1+c\over c}.$$

Let us estimate  $E(T+2U+V) =E(T)+2E(U)+E(V)$.\\
The random variables $T$ and $U$ are the sum of indicator variables:
$$T=\sum_{j=1}^n T_j,$$
  where $T_j(A)=1$ iff the $j$th column of $A$ is all-zero;
$$U=\sum_{1\le i_1<i_2<i_3\le n}U_{i_1, i_2, i_3},$$
where $U_{i_1, i_2, i_3}(A)=1$ iff there exists $i$ such that
$I_i(A)=\{i_1, i_2,i_3\}$, and $W_i(A)=\{1,1,1\}$.\\
Now,
$\displaystyle Pr(T_j=1)=\left ( {{n-1\choose 3}\over {n\choose
3}}\right ) ^L,$
hence
\begin{equation}\label{E(U)}
E(T)\ge n\exp(-3c) (1 -o(1))
\end{equation}
and\\
$\displaystyle Pr(U_{i_1, i_2, i_3}) =L {{n-3\choose 3}^{L-1}\over
{n\choose 3}^L},$ therefore,
\begin{equation}\label{E(V)}
E(U)\ge cn\exp(-9c)(1 -o(1)).
\end{equation}

Finally, let us estimate $E(V)$. In the same way
$$E(V)=\sum_{ 1\le i_1< i_2\le n}E(V_{i_1, i_2}),$$
where $V_{i_1, i_2}(A)=1$
iff there exists $i$ such that $\{i_1,i_2\}\subset I_i(A)$,
$\sum_{k=1}^L a_{k,i_1}=1$,
 $\sum_{k=1}^L a_{k,i_2}=1$, and
$W_i(A)=\{1,1,\alpha\}\hbox{ with }\alpha\ge 2$. \\
Let us introduce the random variables
$Z_{i_1, i_2}(A)$ such that
$Z_{i_1, i_2}(A)=1$
iff there exists $i$ such that $\{i_1,i_2\}\subset I_i(A)$,
$\sum_{k=1}^L a_{k,i_1}=1$ and
 $\sum_{k=1}^L a_{k,i_2}=1$.

\begin{eqnarray*}
Pr(Z_{i_1, i_2}=1)
& = &
Pr(Z_{i_1, i_2}=1\hbox{ and } V_{i_1, i_2}=1)+
Pr(Z_{i_1, i_2}=1\hbox{ and } V_{i_1, i_2}=0)\\
& = &
Pr( V_{i_1, i_2}=1)+
\sum_{i_3\notin\{i_1, i_2\}}Pr(U_{i_1, i_2, i_3}=1)\\
&=&
E(V_{i_1, i_2}) +(n-2) Pr(U_{i_1,i_2, i_3}=1).
\end{eqnarray*}
But,
$$E(Z_{i_1, i_2}) =L(n-2){{n-2\choose 3}^{L-1}\over {n\choose 3}^L}.$$
Therefore,
$$E(V_{i_1, i_2}) =L(n-2){{n-2\choose 3}^{L-1}- {n-3\choose 3}^{L-1}
\over {n\choose 3}^L}.$$
And finally,
\begin{equation}\label{E(W)}
E(V)\ge (3cn\exp(-6c)- 3cn\exp(-9c))(1 -o(1)).
\end{equation}
Finally, Theorem \ref{upper bound thm} follows from (\ref{E(U)}),
(\ref{E(V)}) and (\ref{E(W)}).
\end{proof}

\medskip

The proof of Theorem \ref{upper bound thm} relies on a construction
which can be carried out for all
integer $k\geq 3$. For more readability we have restricted our attention
to the generic case
$k=3$, however  with similar considerations one can obtain the following general result :

\begin{theorem}\label{general upper bound thm} for $k\geq 4:$
$$P_{(L)}\le 1-{\exp(-ck) + c u_k(c) + v_k(c)-1+c \over c} +o(1),$$
where
$$u_k(c)= \sum_{j=1}^{k-3}{k\choose j}\exp(-ck(k-j))((k-j+1)-(k-j-2)(k-j),$$

$$v_k(c)= 3{k \choose 2} \exp(-2ck) + (-k^2 + 3k -1)\exp(-ck^2).$$
\end{theorem}
Therefore, solving  the equation
$$\exp(-ck) + c u_k(c) + v_k(c)-1+c =0,$$
in using MAPLE provides

\begin{equation}\label{upper bound}
{\bf \beta_4\  <\  0.9721, }\quad  {\bf \beta_5\  <\  0.9914},
\quad {\bf \beta_6\  <\  0.9971}
\end{equation}

\section{Experiments}\label{experiments section}

\begin{figure}[h]
\begin{center}
\hspace{-1cm}
\includegraphics[width=100mm,height=62mm]{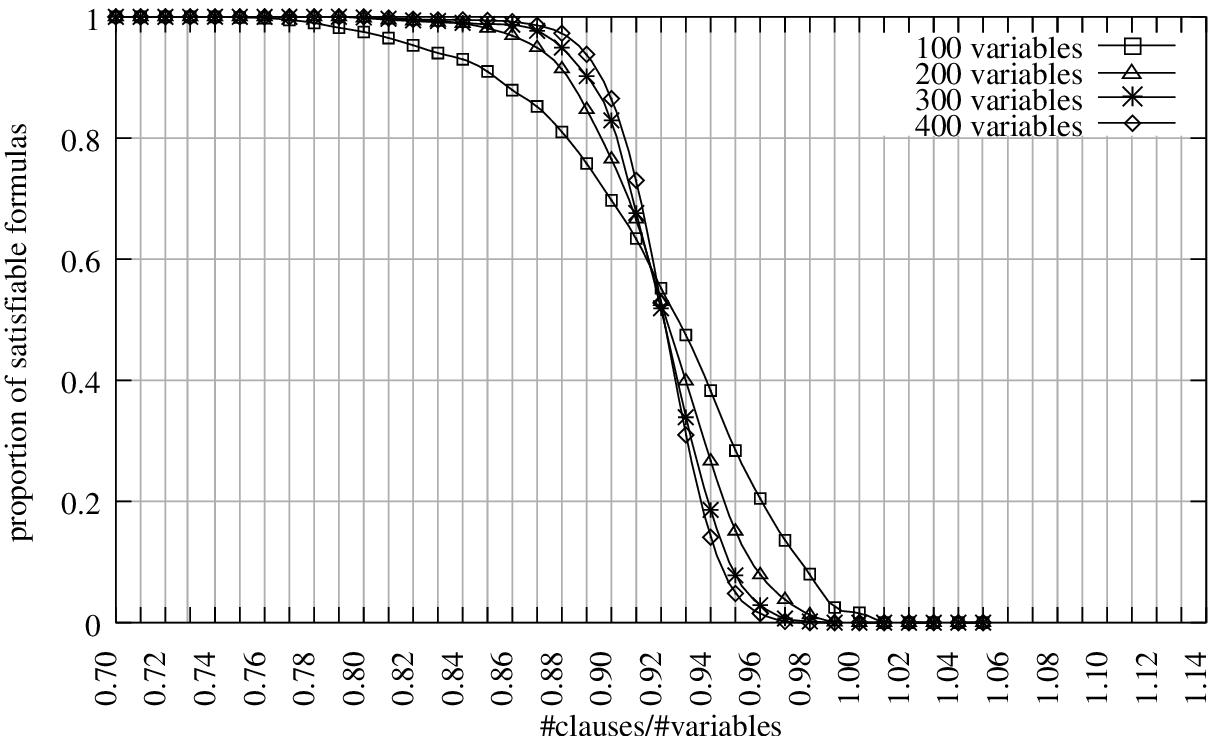}
%\end{center}
\hspace{3cm}\parbox{10cm}{\vspace{-0cm}\caption{}}
\label{fig:ratio_heuristique B3}
\end{center}
\end{figure}

%\vspace{1cm}

%\begin{figure}[h]
%\begin{center}
%\hspace{-1cm}
%\includegraphics[width=100mm,height=62mm]{seuil.new_C2.eps}
%\end{center}
%\hspace{3cm}\parbox{10cm}{\vspace{-0cm}\caption{}}
%\label{fig:ratio_heuristique B3}
%\end{center}
%\end{figure}
 In this section, focussing on the generic case
3-XOR-SAT, we supplement the preceding rigorous results with
experimental results. In the previous sections, we have determined
an approximate scaling of a window in which a phase transition of
the satisfiability must be observed, namely for a ratio number $L$
of clauses to number $n$ of variables below 0.8894, the
probability of satisfiability tends to 1 as $L$ and $n$ tend to
infinity and for a ratio above 0.9278, that probability tends to
0. In order to illustrate the phase transition and to estimate
empirically the location of the critical value $c_3$ of the ratio
for which the transition occurs, we have made experiments
consisting in generating at random (in drawing uniformly and
independently) 3-XOR-SAT formulas over 100, 200, 300 and 400
variables with a ratio varying from 0.70 to 1.14 in steps of 0.1.
For each of these values of ratio, a sample of 1000 formulas has
been solved with a computer program. The proportion of satisfiable
formulas for each considered value of ratio has been plotted on
the above Figure 1. It can be seen that the four smooth lines
connecting the consecutive points corresponding to 100, 200, 300
and 400 variables, straighten as the number of variables increases
showing thus strong empirical evidence of the sharp phase
transition proved in \cite{CreignouD-00}. The crossing of these
lines suggest that the critical value $c_3$ of the transition is a
little lower than 0.92 for 3-XOR-SAT. \vspace{-0.3cm}
\section{Conclusion}
The $\kxorsat$  problem is polynomial time solvable. Compared to
the studies carried out on the phase transition of the SAT
problem, this gives hope to get here an easier study. In a first
step we have made precise the link between the $\kxorsat$'s phase
transition and the rank of sparse random Boolean matrices. This
last problem has been extensively studied by Russian
mathematicians (see \cite{Kolchin-94} and \cite{Kolchin-99}), but
reveals hard combinatorial and probabilistic problems. Our
approach, which consists in using the sharpness of the threshold
in order to specify its location, enables us to get good and
interesting bounds for the critical ratio of  $\kxorsat$'s phase
transition. Indeed, experiments show the tightness of the upper
bound. Moreover let us emphasize that Theorem \ref{lower bound
thm} provides a new proof of Kolchin and Khokhlov's results
\cite{KolchinK-95}, as well as a simpler expression for the
corresponding critical value.

In conclusion, our work  illustrates the importance of directing a
lot of work towards obtaining general conditions for sharpness of
a phase transition.

\end{document}